\documentclass[11pt,a4paper]{article}
\usepackage{jheppub}

\usepackage{verbatim}
\usepackage{bigints}

\newcommand{\ms}{\overline{MS}}

\title{Quasi parton distributions and the gradient flow}
\author[a]{Christopher Monahan}
\author[b,c]{Kostas Orginos}
\affiliation[a]{New High Energy Theory Center and Department of Physics and 
Astronomy, Rutgers, the State University of New Jersey,
136 Frelinghuysen Road, Piscataway, NJ 08854-8019, USA}
\affiliation[b]{Physics Department, College of William and Mary,
Williamsburg, Virginia 23187, U.S.A.}
\affiliation[c]{Thomas Jefferson National Accelerator Facility, Newport News, 
Virginia 23606, U.S.A.}
\emailAdd{chris.monahan@rutgers.edu}
\abstract{We propose a new approach to 
determining quasi parton distribution functions (PDFs) from lattice quantum 
chromodynamics.
By incorporating the gradient flow, this method guarantees that the lattice 
quasi PDFs are finite in the continuum 
limit and evades the thorny, and as yet
unresolved, issue of the renormalization of quasi PDFs on the lattice. In the 
limit that the 
flow time is much smaller than the length scale set by the nucleon momentum, 
the moments of the smeared quasi PDF are 
proportional to those of the light-front PDF.
We use this 
relation to derive evolution equations for the matching kernel that relates the 
smeared quasi PDF and the light-front PDF.}
\arxivnumber{1234.5678}

\begin{document}
\maketitle

\section{Introduction}

Quantum chromodynamics (QCD) is the theory of the
strong nuclear force that connects hadronic bound states to their partonic 
constituents, quarks and gluons. Although quarks and gluons cannot be 
directly accessed in experiments, the connection between hadrons 
and partons can be characterized through parton 
distribution functions (PDFs).

PDFs capture aspects of hadron structure associated with the momentum, 
angular momentum and spin of the constituent quarks and gluons, and play a 
central role in our understanding of high energy hadronic scattering processes  
(see, for example, 
\cite{Brock:1993sz,Sterman:1994ce,Collins:2011zzd}). Through factorization, the 
scattering 
amplitudes of simple scattering processes, such as deep inelastic scattering 
and Drell-Yan production, can be expressed as the convolution of perturbative 
coefficients and PDFs, which encapsulate the nonperturbative dynamics of QCD at 
hadronic scales.

In principle, the direct calculation of PDFs from QCD will provide new insight 
into hadronic structure, more stringent tests of QCD, and reduced systematic 
uncertainties in high energy scattering experiments. At present, however, the 
only systematic method for ab initio, nonperturbative QCD calculations is 
lattice QCD, in which QCD is formulated on a discrete Euclidean hypercube. PDFs 
are defined in terms of matrix elements of light-front wave functions and cannot 
be directly determined from Euclidean lattice QCD. PDFs are currently
determined from global analyses of a wide range of scattering data (see, for 
example, 
\cite{Arbabifar:2013tma,Ball:2013lla,Jimenez-Delgado:2013boa,
Jimenez-Delgado:2014xza,Dulat:2015mca, 
Hou:2016sho,Shahri:2016uzl} for a selection of recent analyses).

Lattice QCD calculations have instead focused on the first 
few Mellin moments of PDFs, which can be related to matrix elements of local 
twist-two operators, where twist is the dimension minus the spin of the 
operator. The lattice 
regulator breaks rotational symmetry, which induces mixing between lattice 
operators that would not mix in the continuum. The mixing between twist-two 
operators of different 
mass dimension introduces power-divergent mixing on the lattice, preventing the 
extraction of more than three moments of PDFs 
\cite{Detmold:2001dv,Detmold:2003rq}.

Recently a new approach to determining PDFs on the lattice was proposed, via 
Euclidean counterparts of PDFs generally referred to as quasi PDFs  
\cite{Ji:2013dva,Ji:2014gla,Lin:2014zya,Gamberg:2014zwa,Ji:2015jwa,
Alexandrou:2015rja,Chen:2016utp}. A similar framework was proposed in 
\cite{Ma:2014jla}. The 
quasi PDFs are Euclidean matrix elements determined at finite nucleon 
momentum. At large Euclidean momentum, the quasi PDFs can be related to the 
true PDFs through an effective theory expansion, the Large Momentum Effective 
Theory (LaMET).

Preliminary lattice calculations have been encouraging 
\cite{Lin:2014zya,Alexandrou:2015rja}, although both 
calculations have incorporated only a single lattice spacing and a full 
understanding of systematic uncertainties is far from complete. In particular, 
there are three challenges for the approach as it stands: the 
restriction to low nucleon momentum with the computational resources currently 
available; a full understanding of the renormalisation of extended Euclidean 
operators; and the precise relation between light-front PDFs and Euclidean quasi 
PDFs. 

These difficulties can be broadly classified as either chiefly practical, or 
chiefly theoretical. The first challenge, that associated with the systematic 
uncertainties corresponding to low values of nucleon momentum on current 
lattices, is largely a practical issue. Studies in the spectator 
di-quark model \cite{Gamberg:2014zwa} suggest that moderate improvements in 
computational resources, and new algorithms tailored to nucleons with large 
momentum \cite{Bali:2016lva}, 
will likely solve this difficulty, at least to a precision that can contribute 
to global analyses of the PDFs in regions of parameter space that are 
experimentally inaccessible. We will not consider these practical 
difficulties any further and focus instead on the theoretical 
aspects of quasi PDFs.

We address one of the theoretical challenges by proposing a new approach to 
calculating quasi PDFs on the lattice, in which the lattice degrees of freedom 
are smeared via the gradient flow 
\cite{Narayanan:2006rf,Luscher:2011bx,Luscher:2013cpa}. Using ringed 
fermions, which do not require any multiplicative wavefunction renormalization
\cite{Makino:2014taa,Hieda:2016lly}, the corresponding lattice matrix elements 
remain finite in the continuum limit. This approach evades the problem of the 
power-divergence associated with the Wilson line operator that defines the 
quasi PDF. The renormalization of quasi PDFs has been viewed through the lens 
of heavy quark effective theory \cite{Ji:2015jwa} and, more recently, a 
counterterm procedure has been proposed to remove this power-divergence 
\cite{Chen:2016fxx,Ishikawa:2016znu}, but neither approach has been established 
beyond two loops in perturbation theory.

Here we examine the relation between the smeared quasi 
PDF and the light-front PDF, and focus on the limit in which the flow 
time is small compared to the 
length scale set by the nucleon momentum and the nucleon momentum is 
sufficiently large that higher twist effects can be neglected. In this limit, 
we express the moments of the smeared quasi PDF in terms of moments of the 
light-front PDF via a small flow-time expansion.
The primary advantage of 
our approach is the finite continuum limit for nonperturbative matrix elements 
determined using lattice QCD. The matching between the smeared quasi PDF, 
regulated by the flow time, and the light-front PDF in, say, the $\ms$ 
scheme can be carried out in the continuum and is independent of the details of 
the nonperturbative lattice calculation.

We start by revisiting the definitions of the light-front and quasi PDFs in 
Section \ref{sec:defs}. We then analyze the relation between the 
light-front and quasi PDFs in Section \ref{sec:spdf} and derive an evolution 
equation for the matching kernel in Section \ref{sec:match}. We 
present our summary and conclusions in Section \ref{sec:concs}.

\section{\label{sec:defs}Distribution functions}

Throughout this work we focus on flavor non-singlet unpolarized quasi and 
light-front PDFs. The extension to polarized quasi PDFs is straightforward. The 
flavor singlet case introduces additional mixing with the gluon distribution, 
but the principles are similar. We also assume that the
quarks are massless and ignore complications arising from the correct 
treatment of heavy flavors, a subject of continued study for light-front 
PDFs (for reviews, see, for example, 
\cite{Thorne:2008xf,Olness:2008px,Binoth:2010nha}).

\subsection{Bare PDFs}

In the following section, when we use the term ``bare'' we mean finite
matrix elements determined with some regulator at finite cutoff. We leave the 
regulator implicit in this discussion, although one can have in mind 
dimensional regularization if desired. These bare matrix elements require 
renormalization in some scheme before one can remove the regulator (or, on the 
lattice, take the continuum limit). This usage follows that of the extensive 
discussions of light-front PDFs in, for example, 
\cite{Collins:1981uw,Collins:2011zzd}.

We denote bare light-front PDFs by 
$f^{(0)}(\xi)$. Light-front PDFs are frequently represented by 
$f_{j/N}^{(0)}(\xi)$, where $j$ 
denotes the quark flavor and $N$ the nucleon 
species, but here we will be considering only non-singlet distributions, for 
which we can neglect mixing between parton species, and work with sufficient 
generality that the nucleon species is not relevant to our discussion. We 
use light-front coordinates, 
$(x^+,x^-,\boldsymbol{x}_\mathrm{T})$ such that $x^\pm = (t\pm z)/\sqrt{2}$, 
and define $\xi=k^+/P^+$. We use $\xi$ to distinguish this variable from the 
Bjorken-$x$ parameter that characterizes the kinematics of scattering 
experiments and is given in 
terms of the experimental momentum transfer $Q^2=-q^2$ and hadron momentum $P$ 
by $x=Q^2/(2P\cdot q)$. The bare PDF is defined as \cite{Collins:2011zzd}
\begin{equation}
f^{(0)}(\xi) = \int_{-\infty}^\infty 
\frac{\mathrm{d}\omega^-}{4\pi}
e^{-i\xi P^+\omega^-} \left\langle P \left| 
T\,\overline{\psi}(0,\omega^-,\mathbf{0}_\mathrm{T})
W(\omega^-,0)\gamma^+\frac{\lambda^a}{2}\psi(0) \right| 
P\right\rangle_\mathrm{C}.
\end{equation}
Here $T$ is the 
time-ordering operator, $\psi$ is a quark 
field, and the subscript C indicates that the vacuum 
expectation value has been subtracted (in other words, only connected 
contributions are included). The operator $W(\omega^-,0)$ is the Wilson line,
\begin{equation}
W(\omega^-,0) = 
{\cal P}\exp\left[-ig_0\int_0^{\omega^-}\mathrm{d}y^-A^+_\alpha(0,y^-,
\mathbf{0}_{\mathrm{T}})T_\alpha\right],
\end{equation}
with ${\cal P}$ the path-ordering operator, $g_0$ the QCD bare coupling, 
and $A^\mu = A^\mu_\alpha T_\alpha$ the $SU(3)$ gauge potential with generator 
$T_\alpha$ (summation over color index $\alpha$ is implicit). The target 
state, $|P\rangle$, is a spin-averaged, exact momentum eigenstate with 
relativistic normalization
\begin{equation}
\langle P' | P \rangle = 
(2\pi)^32P^+\delta\left(P^+-P^{\prime\,+}\right)\delta^{(2)}
\left(\mathbf{P}_\mathrm{T} - \mathbf{P}'_\mathrm{T}\right).
\end{equation}

We define the moments of bare PDFs as
\begin{equation}
a_{0}^{(n)} = \int_0^1 \mathrm{d}\xi\,\xi^{n-1}\left[
f^{(0)}(\xi)+(-1)^n\overline{f}^{(0)}(\xi)\right] = \int_{-1}^1 
\mathrm{d}\xi\,\xi^{n-1}
f(\xi),
\end{equation}
where $\overline{f}^{(0)}(\xi)$ is the anti-quark PDF and the second equality 
follows from the relation of the quark to 
anti-quark PDFs
\begin{equation}\label{eq:pdf2apdf}
f^{(0)}(-\xi) = -\overline{f}^{(0)}(\xi),
\end{equation}
which holds for the bare distributions if the quark and anti-quarks fields are 
classical, or quantized using light-front quantization \cite{Collins:1981uw}.

We can relate these bare moments, $a_{0}^{(n)}$, to matrix elements of 
twist-two operators via
\begin{equation}\label{eq:opePDF}
\left\langle P | {\cal O}_{0}^{\{\mu_1\ldots \mu_n\}} | P 
\right\rangle = 2a_{0}^{(n)} \left(P^{\mu_1}\cdots P^{\mu_n} - 
\mathrm{traces}\right).
\end{equation}
Here the bare
twist-two operators are 
\begin{equation}\label{eq:baretwist2def}
{\cal 
O}_0^{\{\mu_1\cdots \mu_n\}} = i^{n-1}\overline{\psi}(0)
\gamma^{\{\mu_1}D^{\mu_2}\cdots 
D^{\mu_n\}}\frac{\lambda^a}{2}\psi(0)-\operatorname{traces}.
\end{equation}
In these expressions the braces denote symmetrization, $D^\mu$ is the 
symmetric covariant 
derivative, $\lambda^a$ are $SU(2)$ flavor matrices, and the subtraction of the 
trace terms ensures that the operator 
transforms irreducibly under $SU(2)_{\mathrm{L}}\otimes SU(2)_{\mathrm{R}}$. 

\subsection{Renormalized PDFs}

To this point we have considered the bare light-front PDFs, with the 
understanding that such objects are evaluated with some regulator that renders 
the bare distributions finite. We now introduce
renormalized  light-front PDFs. We stress that in this section we 
consider a renormalization scheme that respects rotational symmetry and, for 
definiteness, one can have in mind the $\ms$ scheme.  Complications will arise 
if  a regulator that breaks rotational invariance, such as 
the lattice regulator, is used. We do not discuss such complications here, 
because we will avoid explicit computations of moments 
at finite lattice spacing. All correlation functions computed on the lattice 
can be renormalized and extrapolated to the continuum limit, provided that  
no power divergent mixing exists. In the next section, we propose a smeared 
correlation function that does not have power-divergent mixing.

In general, renormalized light-front PDFs are written in terms of a 
kernel, ${\cal Z}(\zeta/\xi,\mu)$, as
\begin{equation}
f(\xi,\mu) = \int_\xi^1 \frac{\mathrm{d}\zeta}{\zeta}{\cal 
Z}\left(\frac{\zeta}{\xi},\mu\right)f^{(0)}(\zeta),
\end{equation}
where $\mu$ is some renormalization 
scale. We do not need to consider mixing between parton species for 
non-singlet distributions. In terms of the renormalized light-front PDF, the 
renormalized Mellin 
moments are
\begin{equation}
a^{(n)}(\mu) = \int_0^1 \mathrm{d}\xi\,\xi^{n-1}\left[
f(\xi,\mu)+(-1)^n\overline{f}(\xi,\mu)\right] = \int_{-1}^1 
\mathrm{d}\xi\,\xi^{n-1}
f(\xi,\mu),
\end{equation}
which can be related to matrix elements of renormalized twist-two operators, 
${\cal O}^{\{\nu_1\ldots \nu_n\}}(\mu) =Z_{\cal O}(\mu){\cal 
O}_0^{\{\nu_1\ldots \nu_n\}}$, via
\begin{equation}\label{eq:RopePDF}
\left\langle P | {\cal O}^{\{\nu_1\ldots \nu_n\}}(\mu) | P 
\right\rangle = 2a^{(n)}(\mu) \left(P^{\nu_1}\cdots P^{\nu_n} - 
\mathrm{traces}\right).
\end{equation}
This relation holds provided the light-front PDFs and twist-two operators are 
renormalized in the same scheme \cite{Collins:1981uw}.

\subsection{Smeared  quasi PDFs}

We construct a finite quasi PDF matrix element by smearing both the fermion and 
gauge fields via the gradient flow 
\cite{Narayanan:2006rf,Luscher:2011bx,Luscher:2013cpa}. The gradient flow is a 
deterministic 
evolution of the original quark and gluon fields in a new dimension, 
generally referred to as the ``flow time'', towards a classical minimum of the 
QCD action \cite{Luscher:2011bx,Luscher:2013cpa}. The flow-time evolution is 
chosen to remove ultraviolet fluctuations, which corresponds to smearing out 
the quark and gluon fields in real space, with a smearing scale that is 
proportional to the square-root of the flow time. Here we will not describe in 
detail the gradient flow, but refer the reader to the recent reviews 
\cite{Luscher:2013vga,Ramos:2015dla} for more details and applications. 

For our purposes, it is sufficient that the gradient flow has the following 
properties. First, the gradient flow serves as a gauge-invariant ultraviolet 
regulator. Second, given a renormalized theory at zero flow time, the matrix 
elements of smeared fields are automatically finite, up to a multiplicative 
wave-function renormalization for the fermion fields \cite{Luscher:2013cpa}, 
which can be removed by introducing ringed fermion fields 
\cite{Makino:2014taa,Hieda:2016lly}. 
Third, the lattice matrix elements of smeared fields
remain finite in the continuum limit, provided the flow time is fixed in 
physical units \cite{Luscher:2013cpa,Monahan:2015lha}. In essence, the gradient 
flow allows 
one to replace the lattice regulator with a new smearing-scale regulator. This 
last fact allows one to determine the continuum limit of lattice matrix 
elements 
of, for example, twist-two operators, without power-divergent mixing. In the 
continuum, because the gradient flow respects rotational symmetry, the mixing 
between twist-two operators is then reduced to ordinary mixing with 
coefficients 
that depend on the smearing scale and not powers of the inverse lattice spacing 
\cite{Monahan:2015lha}.

We denote the ringed fermion fields at flow time $\tau$ by
$\overline{\chi}(x;\tau)$ and $\chi(x;\tau)$, and the corresponding Wilson line 
at the same flow time, constructed from the smeared gauge fields 
$B_\mu(x;\tau)$, by ${\bf\cal W}(0,z;\tau)$.
We start with the matrix element
\begin{equation}\label{eq:hdef}
h^{(s)}\left(\frac{z}{\sqrt{\tau}},\sqrt{\tau}P_z,
\sqrt{\tau}\Lambda_{\mathrm{QCD}},\sqrt{\tau}M_\mathrm{N}\right) = 
\frac{1}{2P_z}\left\langle 
P_z \left| \overline{\chi}(z;\tau) 
{\bf\cal W}(0,z;\tau)\gamma_z\frac{\lambda^a}{2}\chi(0;\tau)\right| 
P_z\right\rangle_\mathrm{C},
\end{equation}
which, being dimensionless, depends only on dimensionless combinations of 
scales. We note that the flow time has units of length-squared. The subscript C 
indicates that disconnected contributions to this 
matrix element have been removed. The ringed fermion fields require no wave
function renormalization and this smeared matrix element is finite 
provided the flow time, $\tau$, is non-zero and fixed in physical units, 
because correlation functions constructed from smeared fields are finite 
\cite{Luscher:2011bx,Luscher:2013cpa}. Note that divergences will appear in the 
limit of vanishing flow time and the matrix element will then require
renormalization.

We then define the quasi PDF 
\cite{Ji:2013dva,Ji:2014gla} as
\begin{equation}\label{eq:qpdfdef}
q^{\,(s)}\left(\xi,\sqrt{\tau}P_z,\sqrt{\tau}\Lambda_{\mathrm{QCD}}, 
\sqrt{\tau}M_\mathrm{N}\right) 
= 
\int_{-\infty}^\infty \frac{\mathrm{d}z}{2\pi} e^{i\xi z 
P_z} P_z\, 
h^{(s)}(\sqrt{\tau}z,\sqrt{\tau}P_z,\sqrt{\tau}\Lambda_{\mathrm{QCD}}, 
\sqrt{\tau}M_\mathrm{N}),
\end{equation}
where $\xi$ is a dimensionless parameter that can be naively interpreted as the 
longitudinal momentum fraction of the parton in the nucleon $N$.  This 
interpretation is not correct in Euclidean space, however, and instead $\xi$ 
should be viewed as a dimensionless momentum variable in a Fourier 
transformation.  

In practice, the smeared matrix element $h$ is determined from lattice 
computations at finite lattice spacing, $a$, as 
\begin{equation}\label{eq:hacont}
h^{(s)}\left(\frac{z}{\sqrt{\tau}},\sqrt{\tau}P_z,\sqrt{\tau}\Lambda_{\mathrm{
QCD}},
\sqrt { \tau }
M_\mathrm{N}\right) =  \lim_{a\to 0}
h\left(\frac{z}{a},\frac{\sqrt{\tau}}{a},aP_z,a\Lambda_{\mathrm{QCD}},aM_\mathrm
{N}
\right) ,
\end{equation}
where $\sqrt{\tau}$ and $P_z$ are held fixed and
\begin{align}
h\Big(\frac{z}{a},\frac{\sqrt{\tau}}{a},aP_z,a\Lambda_{\mathrm{QCD}},{} & 
aM_\mathrm{N}\Big) = \nonumber\\
{} & \frac{1}{2aP_z}
\left\langle 
aP_z \left| 
\overline{\chi}\left(\frac{z}{a};\frac{\sqrt{\tau}}{a}\right) 
W\left(0,\frac{z}{a};\frac{\sqrt{\tau}}{a}\right)\gamma_z\frac{\lambda^a}{2}
\chi\left(0;\frac { \sqrt{\tau}}{a} \right)\right| 
aP_z\right\rangle_\mathrm{C}.
\end{align}

\section{\label{sec:spdf}Relation to light-front distributions}

We discuss the relation between quasi and light-front PDFs by examining the 
Mellin moments of these distributions, and using the connection between 
Mellin moments and matrix elements of local operators, which are twist-two in 
the case of light-front PDFs \cite{Christ:1972ms}. For the quasi PDFs, the 
local operators corresponding to the Mellin moments do not have a well-defined 
twist, but can be related to twist-two operators after subtracting higher 
twist effects and applying target-mass 
corrections \cite{Lin:2014zya,Alexandrou:2015rja}. Although we 
consider smeared matrix elements in this work, the arguments regarding higher twist and 
target mass effects in \cite{Lin:2014zya,Alexandrou:2015rja} still apply, 
because the flow time serves as an alternative
gauge-invariant regulator to the lattice spacing.

We connect the Mellin moments of the quasi PDF to matrix elements of 
local operators in the following way. Working in axial gauge, $B_z(x;\tau)=0$, 
the 
matrix element $h^{(s)}$ is
\begin{equation}\label{eq:hAzdef}
h^{(s)}\left(\frac{z}{\sqrt{\tau}},\sqrt{\tau}P_z,\sqrt{\tau}\Lambda_{\mathrm{
QCD}},
\sqrt{\tau}
M_\mathrm{N}\right)\Big|_{B_z=0} = 
\frac{1}{2P_z}\left\langle 
P_z \left| \overline{\chi}(z;\tau) 
\gamma_z\frac{\lambda^a}{2}\chi(0;\tau)\right| 
P_z\right\rangle_\mathrm{C}.
\end{equation}
We now substitute this expression into the definition of the quasi PDF, 
Equation \eqref{eq:qpdfdef}, and integrate the resulting
expression over the full range of $\xi$. In contrast to the light-front 
PDF, this range extends from negative to positive infinity, giving
\begin{equation}
\int_{-\infty}^\infty 
\mathrm{d}\xi\, 
q^{\,(s)}\left(\xi,\sqrt{\tau}P_z,\sqrt{\tau}\Lambda_{\mathrm{QCD}},\sqrt{\tau}
M_\mathrm{N}\right)\Big|_{B_z=0} 
= 
h^{(s)}(0,\sqrt{\tau}P_z,\sqrt{\tau}\Lambda_{\mathrm{QCD}},\sqrt{\tau}M_\mathrm{
N}
)\Big|_{B_z=0}.
\end{equation}
Here we have used the relation $\delta(zP_z) = 
\delta(z)/P_z$, for $P_z>0$. We see that the first Mellin moment of the quasi 
PDF can be expressed in terms of the Euclidean matrix element of a local 
(smeared) operator.

We extend this argument to arbitrary 
moments of quasi PDFs by considering derivatives of the quasi distribution with 
respect to the spatial separation $z$ \cite{Collins:2011zzd}. Inverting the 
Fourier transform in Equation \eqref{eq:qpdfdef}, we have
\begin{equation}\label{eq:handqpdf}
h^{(s)}\left(\frac{z}{\sqrt{\tau}},\sqrt{\tau}P_z,\sqrt{\tau}\Lambda_{\mathrm{
QCD}},
\sqrt{\tau}
M_\mathrm{N}\right)
=
\int_{-\infty}^\infty \mathrm{d}\xi\, e^{-i\xi z 
P_z} 
q^{\,(s)}\left(\xi,\sqrt{\tau}P_z,\sqrt{\tau}\Lambda_{\mathrm{QCD}},\sqrt{\tau}
M_\mathrm{N}\right).
\end{equation}
Applying derivatives with respect to the displacement $z$, we obtain
\begin{align}\label{eq:hint1}
\left(\frac{i}{P_z}\frac{\partial}{\partial 
z}\right)^{n-1} 
h^{(s)}{} & 
\left(\frac{z}{\sqrt{\tau}},\sqrt{\tau}P_z,\sqrt{\tau}\Lambda_{\mathrm{QCD}},
\sqrt{
\tau }
M_\mathrm{N}\right) = \nonumber\\
{} & \int_{-\infty}^\infty \mathrm{d}\xi\, \xi^{n-1}e^{-i\xi z 
P_z} 
q^{\,(s)}\left(\xi,\sqrt{\tau}P_z,\sqrt{\tau}\Lambda_{\mathrm{QCD}},\sqrt{\tau}
M_\mathrm{N}\right).
\end{align}

Defining the moments of the smeared quasi PDF, $b_n^{(s)}$, as
\begin{equation}\label{eq:bdef}
b_n^{(s)}\left(\sqrt{\tau}P_z,\frac{\Lambda_{\mathrm{QCD}}}{P_z}, 
\frac{M_\mathrm{N}}{P_z}\right) 
= \int_{-\infty}^\infty 
\mathrm{d}\xi\, \xi^{n-1} 
q^{\,(s)}\left(\xi,\sqrt{\tau}P_z,\sqrt{\tau}\Lambda_{\mathrm{QCD}},\sqrt{\tau}
M_\mathrm{N}\right),
\end{equation}
and substituting the definition of the matrix element $h^{(s)}$, given in 
Equation \eqref{eq:hdef}, into Equation \eqref{eq:hint1}, in the limit that 
$z\rightarrow 0$ we obtain
\begin{align}
b_n^{(s)}\left( \sqrt{\tau}P_z,\frac{\Lambda_{\mathrm{QCD}}}{P_z}, 
\frac{M_\mathrm{N}}{P_z}\right)_{B_z=0}
= {} & \frac{c_n^{(s)}(\sqrt{\tau}P_z)}{2P_z^n} \nonumber\\
{} & \qquad \times \left\langle 
P_z \left| \left[\overline{\chi}(z;\tau) \gamma_z 
\left(i\overleftarrow{\partial}_z^{n-1}\right)\frac{\lambda^a}{2}
\chi(0;\tau)\right ] _{z=0 } \right|P_z\right\rangle_\mathrm{C}.
\end{align}
The perturbative coefficients, $c_n^{(s)}(\sqrt{\tau}P_z)$, capture potential 
singularities in the righthand side of Equation \eqref{eq:hint1} in the limit 
of vanishing separation z and vanishing flow time $\tau$, and follow from a smeared operator product expansion 
\cite{Monahan:2015lha} approach to the nonlocal matrix element, as outlined in 
\cite{Lin:2014zya}.

We restore gauge invariance to obtain our final expression for the moments of 
quasi PDFs in terms of Euclidean matrix elements of local operators:
\begin{equation}
b_n^{(s)} \left(\sqrt{\tau}P_z,\frac{\Lambda_{\mathrm{QCD}}}{P_z}, 
\frac{M_\mathrm{N}}{P_z}\right) 
=\frac{c_n^{(s)}(\sqrt{\tau}P_z)}{2P_z^n}\left\langle 
P_z \left| \left[
\overline{\chi}(z;\tau)\gamma_z (i\overleftarrow{D}_z)^{(n-1)} 
\frac{\lambda^a}{2}\chi(0;\tau) \right ] _{z=0 }
\right|P_z\right\rangle_\mathrm{C} .
\end{equation}

The local operators that appear in the matrix element on the right hand side of 
this expression are not twist-two operators: they are not symmetric and 
traceless. The discrepancy between these matrix elements and matrix elements of 
twist-two operators are given by corrections  that appear in powers of 
$\Lambda_\mathrm{QCD}^2/{P_z^2}$ 
and $M_\mathrm{N}^2/P_z^2$ \cite{Lin:2014zya,Alexandrou:2015rja}. The terms 
that scale as ${\cal O}(M_\mathrm{N}^2/P_z^2)$ correspond to target mass 
corrections \cite{Nachtmann:1973ll,Georgi:1974sr}.  Although the appropriate 
interpretation 
of PDFs in the presence of these target mass corrections is subtle 
\cite{Steffens:2012jx,Monfared:2014nta}, for our purposes it is sufficient that 
these 
non-leading 
corrections can be absorbed 
by writing \cite{Lin:2014zya,Alexandrou:2015rja}
\begin{align}
b_n^{(s)} \left(\sqrt{\tau}P_z,\frac{\Lambda_{\mathrm{QCD}}}{P_z}\right) 
= {} & \frac{c_n^{(s)}(\sqrt{\tau}P_z)}{2P_z^n}\left\langle 
P_z \left|\left[
\overline{\chi}(z;\tau)\gamma_z (i\overleftarrow{D}_z)^{(n-1)} 
\frac{\lambda^a}{2}\chi(0;\tau) \right ] _{z=0 }
\right|P_z\right\rangle_\mathrm{C} \nonumber \\
{} & \qquad \times K_n^{-1}\left(\frac{M_\mathrm{N}^2}{4P_z^2}\right),
\end{align}
where
\begin{equation}
K_n\left(\frac{M_\mathrm{N}^2}{4P_z^2}\right) = \sum_{j=0}^{n/2} 
\bigg(
\begin{array}{c}
n-j \\
j
\end{array}
\bigg)
\left(\frac{M_\mathrm{N}^2}{4P_z^2}\right)^j.
\end{equation}

The corrected matrix 
elements on the right hand side of this equation can now 
be expanded in a Taylor series with respect to
$\Lambda_\mathrm{QCD}^2/P_z^2$. The coefficients in this expansion 
represent higher twist effects that arise because the original matrix 
element is not a matrix element of a twist-two operator. The leading 
coefficient 
in this expansion is a twist-two contribution that can depend only on the 
nucleon structure and the flow time:
\begin{equation}\label{eq:btwist}
b_n^{(s)} \left(\sqrt{\tau}P_z,\sqrt{\tau}\Lambda_{\mathrm{QCD}}\right)  
= 
c^{(s)}_n(\sqrt{\tau}P_z)b_n^{(s,\mathrm{twist}-2)}
\left(\sqrt{\tau}\Lambda_{\mathrm{
QCD}}\right)  +{\cal 
O}\left(\frac{\Lambda_\mathrm{QCD}^2}{P_z^2}\right),
\end{equation} 
where, for $\Lambda_\mathrm{QCD}^2/P_z^2\ll 1$, the higher twist corrections 
can be ignored.

In summary, we assume that: first, we can correct exactly for target mass 
corrections; and second, we can take the momentum $P_z$ sufficiently large that 
higher 
twist effects are negligible. Then, under these 
assumptions, the moments of the smeared quasi PDFs are dimensionless products 
of perturbative coefficients and pure twist-two matrix elements, which are only 
functions of the 
dimensionless quantity
$\sqrt{\tau} \Lambda_{\mathrm{QCD}}$, that contain
information about the structure of the hadron.

\subsection{\label{sec:renormpdf} Short-distance expansion}

We can now relate the moments of the smeared quasi PDF 
$b_n^{(s,\mathrm{twist}-2)}\big(\sqrt{\tau}\Lambda_{\mathrm{QCD}}\big)$, which 
are local matrix elements of smeared fields, to the renormalized moments of the 
light-front PDFs, by using the properties of the gradient flow that arise from 
a short distance expansion 
\cite{Luscher:2011bx,Suzuki:2013gza,Asakawa:2013laa,Makino:2014taa,
Hieda:2016lly}. 
The exponentially local nature of the smearing procedure 
allows for a short distance expansion of the smeared local operators in terms 
of renormalized operators in some renormalization scheme, such as the 
$\overline{MS}$ scheme. It is straightforward to show that this short 
distance expansion leads to 
\begin{equation}\label{eq:shortdist}
b_n^{(s,\mathrm{twist}-2)}\big(\sqrt{\tau}\Lambda_{\mathrm{QCD}}\big) = 
\widetilde{C}_n^{(0)}(\sqrt{\tau}\mu)a^{(n)}(\mu) + {\cal 
O}(\sqrt{\tau}\Lambda_{\mathrm{QCD}}),
\end{equation}
where $\mu$ is a renormalization scale. The leading order term in this 
expansion, $a^{(n)}(\mu)$, is the matrix element of a renormalized twist-two 
operator with the same gamma matrix and derivative structure as the smeared 
operator that appears in the matrix element on the left hand side. The higher 
order terms arise from higher dimension operators that enter the short distance 
expansion of the smeared matrix element.

We now combine this short-distance expansion with Equation \eqref{eq:btwist} to 
write
\begin{equation}\label{eq:double}
b_n^{(s)}\big(\sqrt{\tau}\Lambda_{\mathrm{QCD}}\big) = 
C_n^{(0)}(\sqrt{\tau}\mu, \sqrt{\tau}P_z)a^{(n)}(\mu) + {\cal 
O}\left(\sqrt{\tau}\Lambda_{\mathrm{QCD}},\frac{\Lambda_\mathrm{QCD}^2}{P_z^2} 
\right),
\end{equation}
Both the leading short distance coefficient function,
$ C_n^{(0)}(\sqrt{\tau}\mu, \sqrt{\tau}P_z)$, and the higher order corrections 
can be 
computed in perturbation theory, so that this approximation can be 
systematically improved. 

For the rest of this discussion, we will assume that we work in a regime in 
which there is a hierarchy of scales given by
\begin{equation}\label{eq:scales}
\Lambda_{\mathrm{QCD}},M_N\ll P_z\ll  \tau^{-1/2},
\end{equation}
so that power corrections and higher-twist effects can be ignored. We also 
assume that target mass corrections have been applied.

To relate the smeared quasi PDF with the light-front 
PDF, we introduce a kernel function, $Z(x,\sqrt{\tau}\mu,\sqrt{\tau}P_z)$,
whose Mellin moments are given by
\begin{equation}\label{eq:cinvz}
 \left[C_n^{(0)}(\sqrt{\tau}\mu, \sqrt{\tau}P_z)\right]^{-1} = 
\int_{-\infty}^{\infty} dx\, x^{n-1}  
Z(x,\sqrt{\tau}\mu, \sqrt{\tau}P_z).
\end{equation}
With this definition, and using the properties of multiplicative convolution, 
we find 
\begin{equation}
f(x,\mu) =  \int_{-\infty}^{\infty} \frac{d\xi}{\xi}\, 
Z\left(\frac{x}{\xi},\sqrt{\tau}\mu, \sqrt{\tau}P_z\right) 
q^{\,(s)}\left(\xi,\sqrt{\tau} \Lambda_{\mathrm{QCD}}\right) + {\cal 
O}(\sqrt{\tau}\Lambda_{\mathrm{QCD}}).
\end{equation}
We introduce the inverse kernel through
\begin{equation}
C_n^{(0)}(\sqrt{\tau}\mu, \sqrt{\tau}P_z) = \int_{-\infty}^{\infty} dx\, 
x^{n-1}  
\widetilde{Z}(x,\sqrt{\tau}\mu, \sqrt{\tau}P_z),
\end{equation}
which leads to
\begin{equation}
q^{\,(s)}\left(x,\sqrt{\tau} \Lambda_{\mathrm{QCD}}, \sqrt{\tau}P_z\right) =  
\int_{-1}^{1} 
\frac{d\xi}{\xi}\, \widetilde{Z}\left(\frac{x}{\xi},\sqrt{\tau}\mu, 
\sqrt{\tau}P_z\right) 
f(\xi,\mu) + 
{\cal O}(\sqrt{\tau}\Lambda_{\mathrm{QCD}})
\end{equation}
Note that all of these relations are only 
valid if
\begin{equation}\label{eq:scales_tau}
\Lambda_{QCD},M_N\ll P_z \ll \tau^{-1/2}.
\end{equation}

The kernel function can be 
computed in continuum perturbation theory, following the methods introduced 
in~\cite{Luscher:2011bx} and the examples 
in~\cite{Ma:2014jla,Ji:2015jwa,Chen:2016fxx,Ishikawa:2016znu}.
Given that those computations are performed in the continuum, 
they are independent of the lattice formulation used to extract the 
smeared quasi-PDFs introduced in this paper and lattice computations of 
smeared quasi-PDFs can be performed with a variety of lattice actions, each of 
which results in the same universal continuum quasi-PDF. Consequently, lattice 
computations are disconnected from the matching procedure of these universal, 
continuum quasi-PDFs to the light-front PDFs.  

In both our approach, as in Ji's approach, 
the large nucleon momentum serves only to suppress higher 
twist contributions. We have, however, introduced a new scale, the (inverse) 
flow time, $\tau^{-1}$,  that serves as a regulator of ultraviolet divergences 
that arise from external composite operators  that define the matrix element. The 
scale $\tau^{-1}$ needs to be large relative to hadronic scales but remains 
finite. These 
requirements for the hierarchy of scales, expressed in Equation 
\eqref{eq:scales_tau}, are no different in nature than the requirements used to 
factor physical cross-sections into PDFs and Wilson coefficients and are 
similar in spirit to the factorization approach proposed in 
\cite{Ma:2014jla,Ishikawa:2016znu}. In this approach, the renormalization 
scale, which plays a similar role as our flow time scale,
and the factorization scale are distinct and separate from the large momentum, 
which suppresses higher twist effects. 

\section{\label{sec:match}DGLAP-like equation for the matching kernel}

Ignoring mixing between quark flavors and gluons (i.e. looking at the 
non-singlet distributions) the renormalized PDFs satisfy a DGLAP equation 
\cite{Gribov:1972rt,Dokshitzer:1977sg,Altarelli:1977zs} that describes their 
scale dependence
\begin{equation}
\mu\frac{\mathrm{d}\, f(x,\mu)}{\mathrm{d}\mu} = \frac{\alpha_s(\mu)}{\pi} 
\bigintsss_x^1 
\frac{\mathrm{d}y}{y} f(y,\mu) P\left(\frac{x}{y}\right).
\end{equation}
Here $P\left(z\right)$ is a function whose moments are given by
\begin{equation}\label{eq:gamman}
\int_0^1 \mathrm{d}x\, x^{n-1} P(x) = \gamma^{(n)},
\end{equation}
where 
\begin{equation}\label{eq:rgan}
\left[\mu\frac{\mathrm{d}\;}{\mathrm{d}\mu} -\frac{\alpha_s(\mu)}{\pi}  
\gamma^{(n)} \right]a^{(n)}(\mu) = 0,
\end{equation}
and $\alpha_s(\mu)$ is the (renormalized) strong coupling constant.
 
Similarly, we can derive a DGLAP-like equation for the 
matching kernel that relates smeared quasi PDFs and light-front PDFs. We start 
from the small distance expansion in Equation \eqref{eq:double}, apply 
the renormalization group operator $\mu \,\mathrm{d}/(\mathrm{d}\mu)$, and use 
Equation \eqref{eq:rgan} to derive a renormalization group equation for the 
short distance coefficient
\begin{equation}
\left[\mu\frac{\mathrm{d}\;}{\mathrm{d}\mu} + 
\frac{\alpha_s(\mu)}{\pi}\gamma^{(n)}\right]C_n^{(0)}(\sqrt{\tau}\mu,\sqrt{\tau}
P_z ) = 0+ 
{\cal 
O}(\sqrt{\tau}\Lambda_{\mathrm{QCD}}),
\end{equation}
and its inverse
\begin{equation}
\left[\mu\frac{\mathrm{d}\;}{\mathrm{d}\mu}-
\frac{\alpha_s(\mu)}{\pi}\gamma^{(n)}\right] 
\left[C_n^{(0)}(\sqrt{\tau}\mu,\sqrt{\tau}
P_z)\right]^{-1} = 0+ {\cal 
O}(\sqrt{\tau}\Lambda_{\mathrm{QCD}}).
\end{equation}

We can obtain a DGLAP-like equation for the matching kernel by substituting 
Equations \eqref{eq:cinvz} and \eqref{eq:gamman} into this renormalization 
group equation, to give
\begin{equation}\label{eq:dglapz1}
 \mu\frac{d\;}{d\mu}\,Z\left(x,\sqrt{\tau}\mu, \sqrt{\tau}P_z\right) =  
\frac{\alpha_s(\mu)}{\pi} \bigintsss_{x}^\infty \frac{dy}{y}  
Z\left(y,\sqrt{\tau}\mu, \sqrt{\tau}P_z\right) P\left(\frac{x}{y}\right),
\end{equation}
and
\begin{equation}
 \mu\frac{d\;}{d\mu}\,\widetilde{Z}\left(x,\sqrt{\tau}\mu, \sqrt{\tau}P_z\right) =  -
\frac{\alpha_s(\mu)}{\pi} \bigintsss_{x}^\infty \frac{dy}{y}  
\widetilde{Z}\left(y,\sqrt{\tau}\mu, \sqrt{\tau}P_z\right) P\left(\frac{x}{y}\right),
\end{equation}
up to corrections of ${\cal O}(\sqrt{\tau}\Lambda_{\mathrm{QCD}})$.

\section{\label{sec:concs}Conclusion}

Parton distribution functions (PDFs) characterize nucleon structure in terms of 
the nucleon's constituent quarks and gluons. These PDFs are defined as matrix 
elements of light-front wave functions and cannot be directly calculated in 
Euclidean lattice QCD. In principle, however, the Mellin moments of PDFs can be 
calculated in lattice QCD, through matrix elements of twist-two operators. 
Unfortunately, these calculations are limited to the first few moments, because 
the hypercubic symmetry of the lattice regulator induces power-divergent mixing 
between twist-two operators of different mass dimension, obscuring the 
continuum limit of matrix elements determined on the lattice. Current 
determinations of PDFs rely on global analyses of data in a wide range of 
experimental channels and a determination of PDFs from first principles is 
lacking.

A new approach to determining PDFs in lattice QCD was recently proposed by Ji 
and subsequently, through a related framework, by Qiu and Ma. 
In this approach, one calculates Euclidean quasi PDFs at large nucleon 
momentum. Two recent lattice calculations 
provided promising results, but several aspects of the approach are yet to be 
fully understood. First, there is the practical issue of the systematic 
uncertainties associated with finite nucleon momenta in lattice calculations. 
This issue is likely to be resolved, to the extent that lattice 
calculations will have sufficient precision that results will provide
useful input into global analyses where experimental data are inadequate, 
with improvements in computational resources and the algorithmic advances 
already underway. Second, there are theoretical issues to be clarified: the 
renormalization of the extended operator that defines the quasi PDFs; and the 
relation between the Euclidean quasi PDF and the light-front PDF, which to date 
had been analyzed through a factorization formula at one loop in 
perturbation theory.

We have addressed the first of these theoretical considerations by introducing 
a quasi PDF 
constructed from fields smeared via the gradient flow. 
We explicitly demonstrated that there is a 
simple relation between the 
Mellin moments of the smeared Euclidean quasi PDF and the 
renormalized Mellin moments of the light-front PDF, 
once nucleon mass corrections 
are incorporated and provided the flow time is small relative to the inverse 
nucleon momentum. Corrections to this relation appear at 
${\cal O}(\Lambda_{\mathrm{QCD}}^2/P_z^2)$, where $P_z$ is the Euclidean 
momentum of the nucleon, and ${\cal O}(\sqrt{\tau} \Lambda_{\mathrm{QCD}})$, 
where 
$\tau$ is the flow time. From this correspondence it follows that, provided 
$\Lambda_{QCD},M_N\ll P_z \ll \tau^{-1/2}$, the quasi PDF and light-front PDF 
can be matched through a convolution relation. 

The chief advantage of our approach is that the gradient flow renders the quasi 
PDF finite in the continuum limit and evades the issues of the renormalization 
of 
the non-local operator that defines 
the quasi PDF on the lattice. The resulting continuum matrix elements are 
independent of the choices of discretized action used to undertake 
lattice QCD calculations and can be 
matched directly to the corresponding light-front PDFs in the $\ms$ scheme 
using 
continuum perturbation theory. Combined with a 
nonperturbative step-scaling procedure, this matching can be carried out at an 
energy sufficiently high that perturbative truncation errors are no longer 
uncontrolled. The nonperturbative implementation of our proposal is work in 
progress.

\acknowledgments
We thank Xiangdong Ji, Herbert Neuberger, Jianwei Qiu, and Christian Weisz 
for enlightening discussions during 
the course of this work. We are 
particularly grateful to Carl Carlson for reading an early draft of this 
manuscript. K.O.~has been supported
by the U.S. Department of Energy through Grant Number DE- FG02-04ER41302, and 
through contract Number DE-AC05-06OR23177, under which JSA operates the Thomas
Jefferson National Accelerator Facility.

\clearpage
\bibliographystyle{apsrev}
\bibliography{qpdf.bib}

\end{document}